\documentclass[12pt,a4paper]{article}
\usepackage[a4paper,margin=1in]{geometry}
\usepackage[dvips]{graphicx}
\usepackage{color}
\usepackage{cite}
\usepackage{hyperref}
\usepackage[cmex10]{amsmath}
\usepackage{amsthm}
\usepackage{amssymb}
\usepackage{algorithmic}
\usepackage{url}
\usepackage{algorithm}
\usepackage{subfigure}
\usepackage{wrapfig}

\title{A Framework to Control Inter-Area Oscillations with Local~Measurement}

\author{Abhilash Patel\\
Electrical Engineering\\
Indian Institute of Technology Delhi\\
Hauz Khas, New Delhi 110016, India.\\
Email: abhilash.patel@ee.iitd.ac.in
}
\date{}
\begin{document}

\maketitle
\thispagestyle{empty}
\pagestyle{empty}

\begin{abstract}
Inter-area oscillations in power system limit of power transfer capability though tie-lines. For stable operation, wide-area power system stabilizers are deployed to provide sufficient damping. However, as the feedback is through a communication network, it brings challenges such as additional communication layer and cybersecurity issues. To address this, a framework for synthesizing remote signal from local measurement as feedback in the wide-area power system stabilizer is proposed. The remote signal is synthesized using different variants of observers in a case study of two-area benchmark system. The proposed framework can improve the damping of inter-area oscillations for static output feedback controller. The presented framework should help to design attack-resilient controller design in smart grid.


\end{abstract}

\section{Introduction}

With an increase in power demand and installation of distributed energy sources in the grid, the stability of power transfer to distant geographical locations  has evolved into a challenging issue. Oscillations in a power system can limit the power transfer capability between different areas. These oscillations are contributed by different modes of system dynamics. One example of such modes is an electromechanical mode, which is characterized by low frequency. In general, these electromechanical modes are poorly damped, raising a threat to system stability under limiting power transfer. These electromechanical modes are further grouped as local modes and inter-area modes based on their participation in the system states. In local modes, synchronous machines in one area oscillate against the synchronous machines from the same area and in inter-area modes, machines from one area oscillate against machines from another area connected through a tie-line~\cite{klein91}. To damp the local modes, local power system stabilizer (PSS) with feedback of local measurements are used. However, as the inter-area modes are relatively less observable in local signals, a combination of remote and local signals are being used as feedback to design a wide-area power system stabilizer~\cite{aboul96}. The objective of a wide-area power system stabilizer is to increase the damping of inter-area modes using the measurement from both the local area as well as a remote area.

The remote signals are usually measured by the phasor measurement unit of the wide-area measurement system in the grid~\cite{phadke08}. These measurements are used for control, monitor and protection purposes~\cite{chakrabortty2013}. However, as the sharing of information is through a network, it can be vulnerable to cyber-attack~\cite{ashok2017,marinovici2019}. It has been noted that such attack in the communication channel can lead to instability in the system or degraded performances.

In this letter, we aim to develop a framework that can be used to damp inter-area oscillations using local measurement. In this scenario, the remote signals are synthesized from local measurement using observer from control theory. The convergence of error can be ensured using Lyapunov based analysis. Such a framework increases the grid resilience towards attack in communication as well as reduces the requirement of abundant sensors.

\section{A Benchmark Smart Grid}

\begin{figure}[!tbh]
\centering
\includegraphics[width=3.5in]{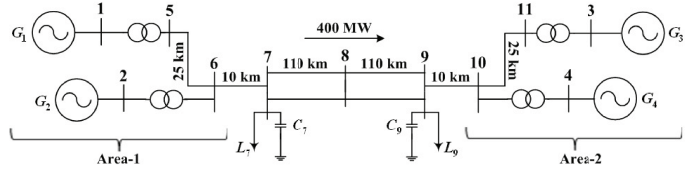}
 \caption{The 4 Machine 11 Bus study system~\cite{kundur94book} }
\label{fig:2_area_system}
\end{figure}

The benchmark two-area system \cite{kundur94book} (Fig. \ref{fig:2_area_system})  is considered in this study. The power system network has 11 buses and 4
generators (Fig. \ref{fig:2_area_system}). Two areas connected by   a weak tie-line to transfer power. Each area of the
system consists of two generators and Local PSSs are installed   at $G_1$ and $G_3$ terminals to provide sufficient damping of local modes. However, it is to be noted that
\cite{aboul96}, such local controllers are insufficient to improve
inter-area damping without a wide-area signal as feedback.
The system is modeled in PST and linearized around an equilibrium point. In general the linearized model of this system can be written as,
\begin{equation}
\label{eq:linsys}
\begin{split}
\dot{x}&=Ax+Bu+Ed,\\
y_l&=C_lx,\\
y_r&=C_rx,\\
\end{split}
\end{equation}
where $x$ is system states, $u$ is the input, $d$ is the disturbance, $y_l$ is the local output and $y_r$ is the remote output. The matrices $A$, $B$, $E$, $C_l$, $C_r$ are of appropriate dimensions.

The low-frequency swing modes are identified with participation analysis \cite{kundur94book}. The modes which have higher participation of electromechanical states $\delta_i,\ \Delta\omega_i$ ($\delta$ is the rotor angle, $\Delta\omega$ is the speed deviation of the synchronous machines) states are the swing modes. The swing modes come out to be $M_1:-0.037\pm j3.90$, $M_2:-1.03\pm j6.8$, $M_3:-0.81\pm j7.2$. The damping of the $M_1$ is found out to be $0.009$, which needs to be improved. It is to be noted that states of $G_1$ and $G_3$ which are farthest from the tie-line have higher participation in the inter-area mode.

To avoid the adverse effect of inter-area damping injection to other modes of the system, the wide-area loop is selected which has higher controllability and observability corresponding to the inter-area mode. This is done by geometrical measures, following \cite{hamdan88}, as:
\begin{equation*}
LSI_{mn}=\frac{\lvert {b_m}^Tv_l\rvert \lvert c_nv_r\rvert}{\|b_m\|\|v_l\|\|c_n\|\|v_r\|}
\end{equation*}
where $v_l$ and $v_r$ are the left and right eigenvector corresponding to the inter-area mode respectively; $b_m$ represents the column vector of $m^{th}$ input; $c_n$ represents the row vector of $n^{th}$ output.
As $\Delta\omega_{24}$ for feedback signal and $G_2$ for input has the highest Loop Selection Index (LSI), it is selected as the wide-area loop.

\section{Remote Signal Synthesis using Observer}

To illustrate the framework, Luenberger observer structure is considered to estimate the states of the system. These estimated states are mapped into the remote signal estimate using the system model. 
The Luenberger observer with output mapping can be written as,
\begin{equation}
\begin{split}
\dot{\hat{x}}&=A\hat{x}+Bu+L(y_l-C_l\hat{x})\\
\hat{y}_r&=C_r\hat{x}
\end{split}
\label{eq:luenobs}
\end{equation}
where $\hat{x}$ is the estimated system, $\hat{y}_r$ is the estimated remote measurement, and $L$ is the observer gain.
The error is defined between the estimated values and actual values as,
\begin{equation}
e(t)=x(t)-\hat{x}(t).
\end{equation}

For the system (\ref{eq:linsys}) and observer (\ref{eq:luenobs}), the error dynamics is
\begin{equation}
\begin{split}
\dot{e}&=(A+LC)e+Ed=A_{cl}e+Ed\\
\end{split}
\label{eq:luenobs}
\end{equation}

The qualitative behavior of the error dynamics can be inferred from Lyapunov based analysis~\cite{khalil2002nonlinear}. Consider a quadratic Lyapunov function,
\begin{equation}
\begin{split}
V(e)=e^TPe
\end{split}
\end{equation}
and the time-derivative of  Lyapunov function is,

\begin{equation}
\begin{split}
\dot{V}(e)&=\dot{e}^TPe+e^TP\dot{e}\\
&=e^T(PA_{cl}+A_{cl}^TP+)e+d^TPE^TEPd\\
&=e^TQe+d^TPE^TEPd\\
&<\lambda_{min}(Q)\|e\|^2+\|EPd\|^2\\
&<\frac{\lambda_{min}(Q)}{\lambda_{max}(P)}V+\|EPd\|^2
\end{split}
\end{equation}

In the absence of disturbance, the error converges to zero exponentially. The convergence rate can be increased by increasing the minimum eigenvalue of $Q$ or decreasing the maximum eigenvalue of $P$. Such results termed as exponential stability of error dynamics~\cite{khalil2002nonlinear}. However, in the presence of a disturbance, the error may not be zero, but the stability of the error dynamics retain in the sense of input-to-state stability.
The observer gain is designed such that $A+LC$ meets Hurwitz criterion. This is to ensure the eigenvalues of $Q$ are negative.

\begin{figure}[!tbh]
\centering
\includegraphics[width=3in]{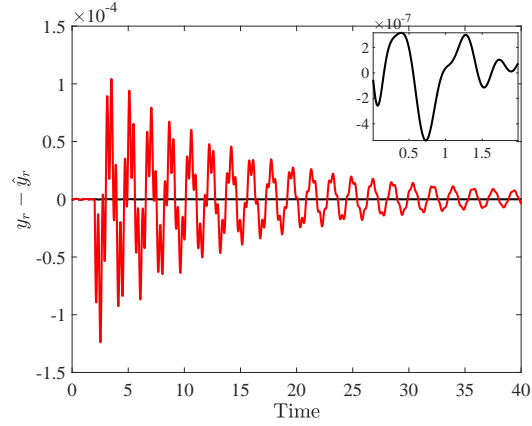}
 \caption{The difference between the estimated remote signal and the actual remote signal. The black line represents the performance without any disturbance and red line represents the performance in presence of disturbance.}
\label{fig:error_knoinp}
\end{figure}

The observer is implemented in the benchmark two-area system. The local measurement $\Delta\omega_2$ is measured and used to estimate the remote measurement of $\Delta\omega_4$. The observer can estimate the remote signal in the absence of disturbance. To verify in simulation, different initial conditions were selected, and error converges to zero as shown in Fig. \ref{fig:error_knoinp}. The disturbance is simulated by changing the mechanical power input as a pulse in the $G_2$ terminal. In the presence of disturbance, the error does not converge to zero. Perhaps, this is due to the lack of information on disturbance to the observer.
This non-zero and prolonged transient in error can deteriorate the performance of wide-area control.

%
%
%
%

\section{Remote Signal Synthesis using Observer in Presence of Disturbance}

To address the shortcoming of Luenberger observer, a different structure of observer is adopted which can undertake disturbance while estimating the states~\cite{chen2012robust}. The observer structure is,
\begin{equation}\begin{split}
\dot{z}&=Fz+TBu+Ky\\
\hat{x}&=z-Hy\\
\hat{y}_r&=C_r\hat{x}
\end{split}\end{equation}

The error dynamics is
\begin{equation}
\begin{split}
\dot{e}&=\dot{x}-\dot{\hat{x}}\\
&=(I+HC)(Ax+Bu+Ed)-(Fz+TBu+Ky)\\
&=(MA-KC)x+(MB-TB)u+MEd-F(Mx-e)\\
&=Fe+(MA-KC-FM)x+(MB-TB)u+MEd
\end{split}
\end{equation}

From the error dynamics, existence conditions can be noted as follows,

\begin{enumerate}
\item $ME=0\implies (I+HC)E=0 \implies H= -E(CE)^{-1}$
\item The pair $(MA,C)$ must be detectable
\item $MB=TB$
\item $MA-KC-FM=0 \implies F= MA-LC \mbox{ where } L=K+FH$
\end{enumerate}

If these existence conditions are satisfied, the error dynamics reduced to,
\begin{equation}
\dot{e}=(MA-LC)e
\end{equation}

The convergence of the error dynamics can be exploited using Lyapunov theory. Consider a Lyapunov function 
\begin{equation}
V(e)=e^TPe
\end{equation}
and 
For the error to be zero in time, $dot{V}(e)<0$, where 

\begin{equation}
\begin{split}
\dot{V}(e)&=e^TPe\\
&=\dot{e}^TPe+e^TP\dot{e}\\
&=e^T(A^T M^T P+PMA-C^TL^TP-PLC)e\\
&=e^TQe.\\
\end{split}
\end{equation}

The estimation error will converge to zero if $A^T M^T P+PMA-C^TL^TP-PLC<0$. However, this inequality is bilinear in nature and difficult to solve. The inequality can be transformed into a linear form by considering $R=PL$. Hence the following equations need to be solved for obtaining the observer gain,

\begin{equation}\begin{split}
&A^T M^T P+PMA-C^TR^T-R^TC<0\\
&L=P^{-1}R.
\end{split}\end{equation}

\begin{figure}[!tbh]
\centering
\includegraphics[width=3in]{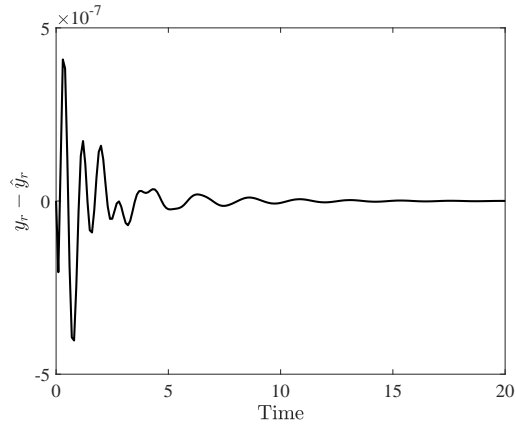}
 \caption{The difference between estimated remote signal and actual remote signal.}
\label{fig:error_unknoinp}
\end{figure}

\begin{figure}[!tbh]
\centering
\includegraphics[width=3in]{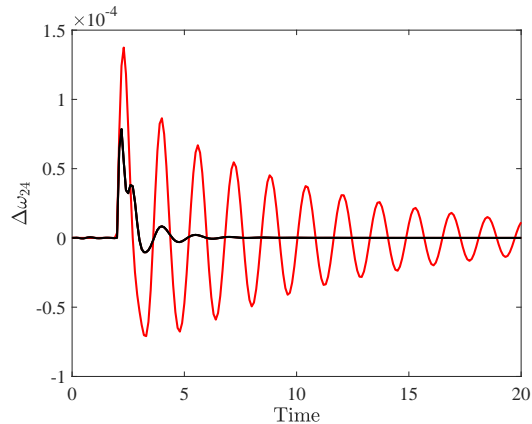}
 \caption{Performance of wide-area power system stabilizer for local signal measurement and synthesized remote signal as feedback. Red line indicates performance without any wide-area controller and black line indicated performance of wide-area controller with synthesized remote signal. }
\label{fig:dw24_unknoinp}
\end{figure}

The extended observer is implemented in the same benchmark two-area system. The local measurement $\Delta\omega_2$ is measured and used to estimate the remote measurement of $\Delta\omega_4$. 
As illustrated in Fig. \ref{fig:error_unknoinp} the observer successfully estimated the remote signal even in the presence of disturbance. The disturbance is simulated by changing the mechanical power input as a pulse in the $G_2$ terminal.  The synthesized remote signal and locally measured signal are fed to a static wide-area controller to validate the ability to damp inter-area oscillations. It can be noted the controller can damp out the oscillations quickly within two cycles.

\section{Conclusion}

In general, to effectively damp inter-area oscillations, a wide-area signal based controller is used. In such a case, remote signals are shared over a communication network. Here, we developed a framework for the smart grid to synthesize the remote signal using local measurement. A standard Luenberger observer lacks the capability to estimate correctly in the presence of disturbance. Hence, another structural observer is applied that can be robust to unknown disturbance. The effectiveness of the proposed framework is validated in a two-area system, showing the controller can damp out inter-area oscillations rapidly with local measurement only.

%
%
%

%

\bibliography{ReferencesWAC}
\bibliographystyle{IEEEtran}
\end{document}